# Integrating Sustainability in Controlling and Accounting Practices: A Critical Review and Implications for Competences in German Vocational Business Education

Julia Pargmann[1,2] and Florian Berding[1]

[1]University of Hamburg, Sedanstraße 19, 20146 Hamburg, Germany. E-Mail: julia.pargmann@uni-hamburg.de; florian.berding@uni-hamburg.de

[2]Corresponding author

**Abstract**

Sustainability in accounting and controlling has traditionally been understood in terms of securing the long-term existence of companies. However, with the introduction of integrated non-financial reporting, sustainability, as per the triple bottom line model, is increasingly being discussed as a component of accounting and controlling. Yet, integration primarily occurs in separate sustainability management and controlling departments. Moreover, the implementation of sustainability efforts requires suitably qualified employees, who drive the transition. The academic discourse surrounding sustainability in businesses in general, and in accounting and controlling specifically, is complex. It remains unclear to what extent sustainability has been integrated into accounting and controlling, and what competencies employees need to manage this transformation. These questions will be critically analyzed in this structured literature review of 79 publications. The results provide insights into a) how companies conceptualize sustainability, b) whether and how they integrate it into their value creation processes, and c) the relevance of accounting and controlling for these developments. To contextualize the role of employees, the competency requirements within companies will be analyzed to enable employees in accounting and controlling to engage effectively in sustainability-oriented activities. Specifically, implications for changes in curricula with a focus on accounting and controlling are derived.

**Author contributions**

Julia Pargmann initiated the project, drafted the theoretical background and methodology, analysed and interpreted the data and was a major contributor in writing the manuscript. Florian Berding served as a sparring partner in the development of the project idea and its methodology and was a contributor in writing and editing the manuscript. All authors read and approved the final manuscript.

**Declarations**

**Funding**

No funds, grants, or other support was received.

**Acknowledgements**

Not applicable.

**Availability of data and materials**

The datasets used and/or analysed during the current study are available from the corresponding author on reasonable request.

**Competing interests**

The authors have no relevant financial or non-financial interests to disclose.

**Introduction**

Against the backdrop of climate change, society, industry, and the economy are called upon to live, produce and do business more sustainably. One area of business that has been discussed in the context of sustainability from many perspectives is the area of accounting and controlling. Generally, controlling processes are used to help with corporate decision-making.

Sustainability in accounting and controlling was previously understood primarily in terms of securing the long-term existence of the company. However, partly due to the introduction of integrated non-financial reporting, sustainability in the sense of the triple bottom line is also increasingly being discussed as a component of accounting and controlling (Meeh-Bunse & Luer, 2016; Osburg, 2013). Currently, this primarily takes place in separate sustainability management and controlling departments instead of an integrative manner (Ghosh et al., 2019; Schaltegger, 2016). The implementation of sustainability efforts also requires suitably qualified employees, so-called change agents, who drive corporate change towards sustainability (e.g., Berding et al., 2020; Bliesner-Steckmann, 2018; Gallagher et al., 2020). In addition to a well-developed action competence, these employees need a special skill set as they are often confronted with rejection and insecurity as they motivate change.

Both accounting and controlling are subjects that play a vital role in vocational business education as they help students to grasp complex economic correlations, effect chains and the consequences of their own economic decisions. This process-oriented thinking style enables students to apply their knowledge to new contexts and challenges. The overall goal of vocational education, the development of action competence, helps students react more professionally under changing conditions. One of these conditions is the necessity to do business sustainably by managing the triple bottom line activities of a company.

To foster the global development towards sustainability, an increasing number of companies produces integrated reports that include environmental, social and governance aspects. In the European Union, legislation requires companies to report on their sustainability actions, 96% of the 250 largest companies worldwide publish sustainability reports (KPMG, 2020). The controlling department plays a vital role in management and control processes, as it collects, analyzes, interprets, and prepares corporate data for leadership decision-making. Regarding the increasing importance of corporate sustainability, it seems logical to include these employees in the developments. However, only few companies include their regular controlling and accounting departments (Schaltegger, 2020). While controlling and accounting has long served to ensure economic longevity, the introduction of non-financial, integrated sustainability reports has started discussions on the role of the controlling department (Osburg, 2013). Sustainability will become relevant for students in business education to keep up with industry standards. However, most curricula in German vocational business education still focus on environmental protection instead of a holistic approach (Fischer et al., 2023).

In vocational business education, accounting and controlling learning materials are usually focused on understanding double-entry bookkeeping and getting to know classic control instruments like strength-weaknesses-opportunities-threats (SWOT) analysis, different KPI such as return on investment (ROI) or ABC analysis. Rarely do they complete the control cycle, leaving out a crucial part that is supposed to teach students how to make economic decisions (Stütz et al., 2022).

In recent years, the number of publications on sustainability in accounting and controlling has drastically increased. Besides sustainable transformation efforts, these publications also discuss the skills that employees need to contribute to these changes. To successfully include sustainability in accounting and controlling classes, teachers need to know which types of content are relevant for their students and in which way these contents are connected to different competence dimensions.

Thus, both the dimension of corporate change towards sustainability and the competences that companies require from their employees to enable this change are areas of interest. This points to a research gap on the integration of sustainability into controlling and accounting education as it is unclear where the academic discourse on corporate sustainability currently is at, and which skills students need to keep up to date. This paper aims to contribute to this gap with a pilot analysis. The aim is to develop a benchmark for further implications on competences. A systematic review of n=79 publications was carried out and evaluated using content analysis (Kuckartz, 2018). Contributions to the following research questions will be made:

RQ1: How is sustainability discussed in accounting and controlling literature? Which tools, processes and practices are relevant in this context?

RQ2: Which sustainability-focused competencies are discussed in accounting and controlling research and practice?

The aim of this article is to derive the sustainability-related professional, methodological and social requirements that students face in their training companies and for which they need to be prepared in classroom activities. We focus on accounting and controlling activities as these topics are often used in vocational educational settings to illustrate complex economic connections and derive consequences of management decisions. Additionally, we collect information on the extent to which employees in accounting and controlling (must) take on the role of change agents and which competence requirements employees will have to meet in the future. As a result of this critical analysis, we extract implications for accounting and controlling curriculum development.

The rest of this paper is structured as follows: First, we introduce central terminology and develop the methodology as well as the categories for the critical analysis. Then we report on the results through a structured content analysis. Lastly, we will discuss the findings regarding the role of sustainability and derive implications for vocational accounting and controlling education to help both research and practice to sharpen their efforts on vocational sustainability education. A comment on limitations and conclusions ends this paper.

**Methodology**

*Terminology*

The World Commission on Environment and Development (1987) operationalizes sustainability to meet "the needs of the present without compromising the ability of future generations to meet their needs". This rather broad definition is commonly conceptualized through the triple bottom line model that integrates the economic, ecological and social dimensions (see Purvis et al. (2019) for an overview of conceptual origins). While there are more recent efforts to shape a more inclusive definition (Virtanen et al., 2020), the triple bottom line approach remains widely known and integrated. Most businesses inherently strive for economic sustainability but are not necessarily inclined to contribute to ecological or social justice. To analyze sustainability within supply chains, three archetypical strategies can be used: sufficiency, consistency, and efficiency (Huber, 1995). Their aim is to increase an institution's sustainability. While the optimization of input vs. output (i.e., material, energy, and other resources), mostly through technology, constitutes efficiency, consistency pertains to increasing the circularity of value streams. The last strategy, sufficiency, means to only consume or produce what is needed to satisfy needs without risking other people's needs satisfaction (Brinken et al., 2022).

*Literature Search and Analysis*

When designing this literature review, the preferred reporting items for systematic reviews and meta-analyses (PRISMA) (Page et al., 2021) were used to ensure a quality analysis. Figure 1 shows the PRISMA flow diagram for this paper.

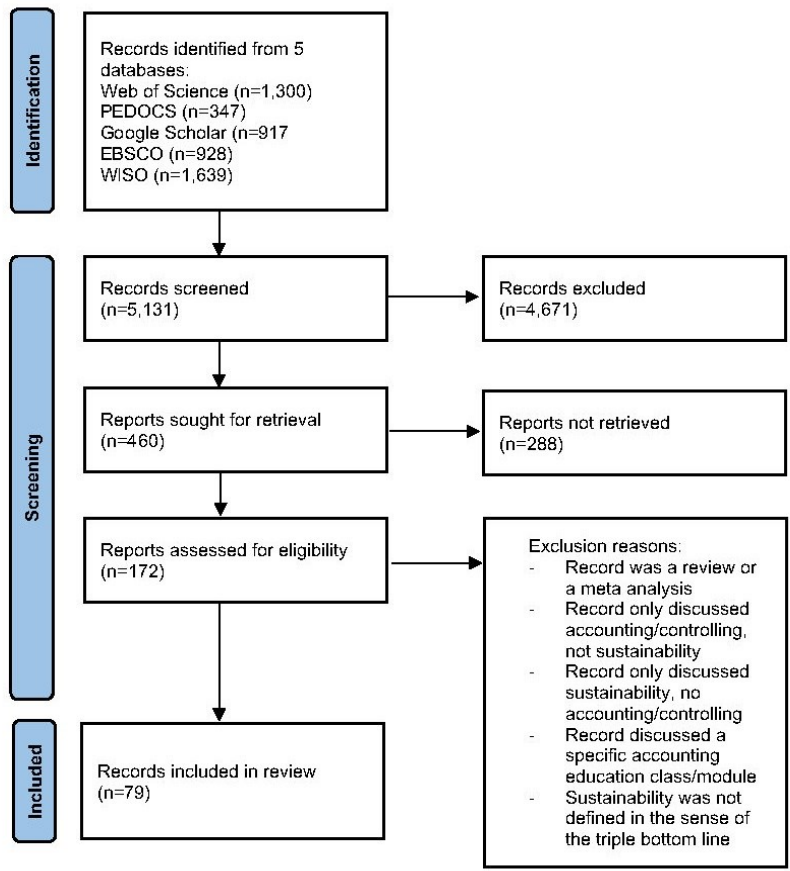

**Fig. 1** PRISMA Flow Diagram

The search strategy included all accounting and controlling publications with a sustainability focus between 2000 and 2022. This yielded 5,131 records for screening of which 4,671 were excluded right away as they did not connect the two perspectives of sustainability and controlling/accounting. Publications about accounting or controlling classes in higher education/on a specific class or module were also excluded, even if they had a sustainability focus. Additionally, the screening excluded all publications which did not use a suitable definition of sustainability, i.e., in the sense of longevity. The review only includes original research or reports from practice, no literature reviews or meta-analyses. The search terms were grouped to search for sustainability and controlling/accounting literature as well as literature that includes competences, skills, or knowledge/expertise in this context. Table 1 presents an overview of keywords used.

**Table 1** List of Keywords

| Group No. | List of Keywords |
|---|---|
| 1 | Sustainability, ecology, sustainable development goals, non-financial reporting, corporate social responsibility, integrated reporting, sustainability controlling, accounting for sustainability, financial accounting, cost accounting, investment accounting, external accounting, sustainable key performance indicators |
| 2 | competence, skills, activity, education, training, learning process, accounting, controlling competence, skills, activity, education, training, professional competence, action competence, methodological competence, social competence, self-competence, accounting, controlling, sustainability controlling |

Because oftentimes, companies use sustainability efforts as PR and the aim of this paper is to identify sustainability efforts (even if optimized for marketing), the search strategy was not limited to scholarly papers but instead included grey literature, management papers and book chapters as well. We derived a total of n=79 publications for coding.

Papers classified as 'relevant' were coded using an analytical framework (Appendix A). After a baseline coding, the remainder was coded in pairs with regular discussion regarding inconsistencies. The coding manual was developed using guidelines for qualitative content analysis (Kuckartz, 2018) and was carried out using the software MaxQDA (Release version 22.8.0).

Most of the analytical framework was developed deductively, as the purpose of this paper is not to create a new systemization but rather to see how far the discourse on sustainability in controlling and accounting has progressed. The following categories were used to enable a critical analysis (see Appendix A for a tabular overview over main and sub-categories):

- **Competence requirements:** In the German-speaking countries, the promotion of action competence is a crucial part of vocational education programs. This concept depicts the ability to solve occupational tasks successfully and to execute routine and non-routine activities. Typically, frameworks distinguish between professional, social and self-competences, with methodological and technological competences as interdisciplinary dimensions (Roth, 1971; Winther and Achtenhagen, 2008). In this review, we are interested in the connection between sustainability and desired competences. Thus, the above concept was expanded by the notion of sustainability in the different competence dimensions.

- **Approach to sustainability:** To deduce where corporate sustainability efforts are located and how well companies have integrated sustainability into their supply chain, the category of business processes was added. The differentiation between controlling, core and support processes helps match sustainability contents with accounting and controlling activities (Gadatsch, 2017). The areas of activities were added because not every company, especially not small- and medium-sized enterprises (SME), are organized in a process-oriented manner (Wöhe et al., 2020). Here, sustainability efforts are closely intertwined between different areas of the business.

- **Sustainability dimensions, strategies, and conflicts:** These categories were included to develop an understanding of how companies conceptualize their sustainability efforts, how they define them and how long-term they think about the topic (Perridon et al., 2017). The most common definition of sustainability is a balance between the ecological, social and economic dimensions (Schmidt, 2012) and a sustainable development requires a transformation that stops the exploitation of resources (World Commission on Environment and Development [WCED], 1987). In a simplified way, sustainability efforts can follow one of three strategies: efficiency, consistency, or sufficiency. Each of these strategies has its own risks and benefits and their implementation depends on the business model. In consequence, there are some conflicts that can be derived from the triple bottom line and the sustainability strategies, i.e., rebound effects. As these also affect controlling and accounting, they were included. We also inductively coded the motivation of companies to become more sustainable.

- **Relevant certificates and frameworks, tools, and methods:** Corporate sustainability efforts usually include some notion of corporate social responsibility and thus integrated reporting. Due to legislative developments, companies registered in the European Union that exceed the size of 500 employees and are of public interest are now obligated to report on their efforts in a suitable way, regulations for other enterprises are to follow in the coming years (Directive (EU) 2022/2464 of the European Parliament and of the Council, 2022). For the purpose of this paper, it is of interest to identify reporting standards and certificates in which businesses have a particular interest in (Scholz and Pastoors, 2018). As we focus on controlling and accounting, the 'tools and methods' category also includes key performance indicators (KPI) used by companies. This is especially interesting because in vocational business education in Germany, the curricula use KPIs as means to illustrate impact relationships to students.

**Results**

*Description of the sample*

Out of the 79 publications 25 were journal articles, 33 were magazines, whitepapers, or grey literature, 3 were conference proceedings, 15 were book chapters and 3 were monographs. Regarding the publishing years, three peaks can be identified in the years 2011 2016 and in 2020/2021. Nineteen publications are from the period between 2001 and 2011, 60 were published between 2012 and 2022. This creates the impression that the implementation of sustainability into controlling and accounting has gained importance as more authors publish about it. There are clear peaks over time (Fig. 2). Language-wise, most publications (n=55) were written in German, the remainder (n=24) are in English.

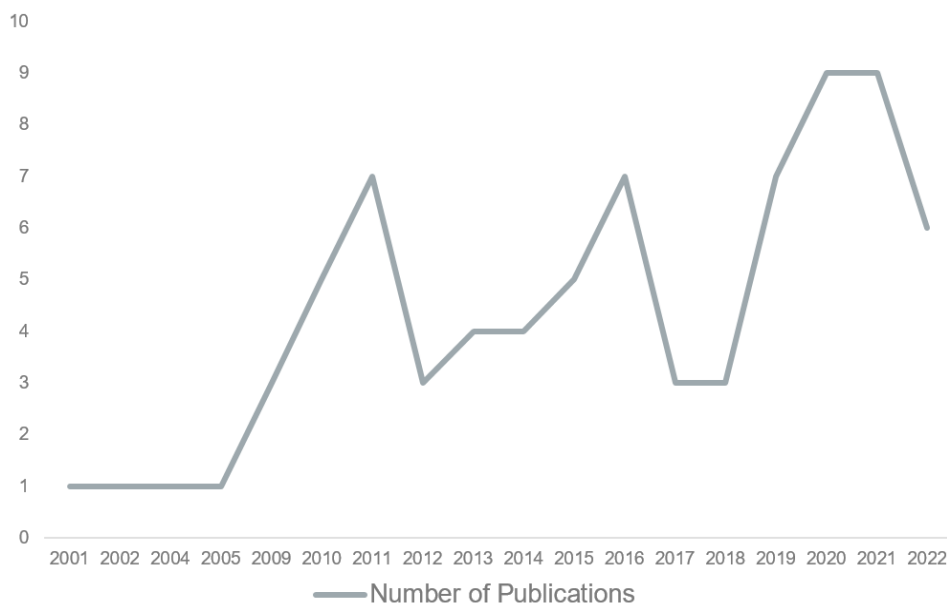

**Fig. 2** Number of publications analyzed.

***RQ1: How is sustainability discussed in accounting and controlling literature? Which tools, processes and practices are relevant in this context?***

*Approaches towards sustainability*

Interestingly, most texts do not focus on the economic dimension of the triple bottom line (Fig. 3, left). Instead, the ecological dimension is discussed most frequently, followed by the social dimension. This finding could be due to two reasons: a) as controlling and accounting are the main topic of all texts, the economic sustainability of a company could be seen as a given, as without it, there would be no reason to enforce social or ecological sustainability; b) the discourse on sustainability has gone beyond the economic dimension as without social and ecological justice, there will be no true economic longevity as global consumer priorities shift. Depending on the argument, both perspectives serve their purpose and seem like viable options when conflicts between the different dimensions are analyzed (Fig. 3, right). Here, conflicts between the social and ecological dimension seem to be less important, as conflicts between economic and ecological decisions and economic and social decisions dominate.

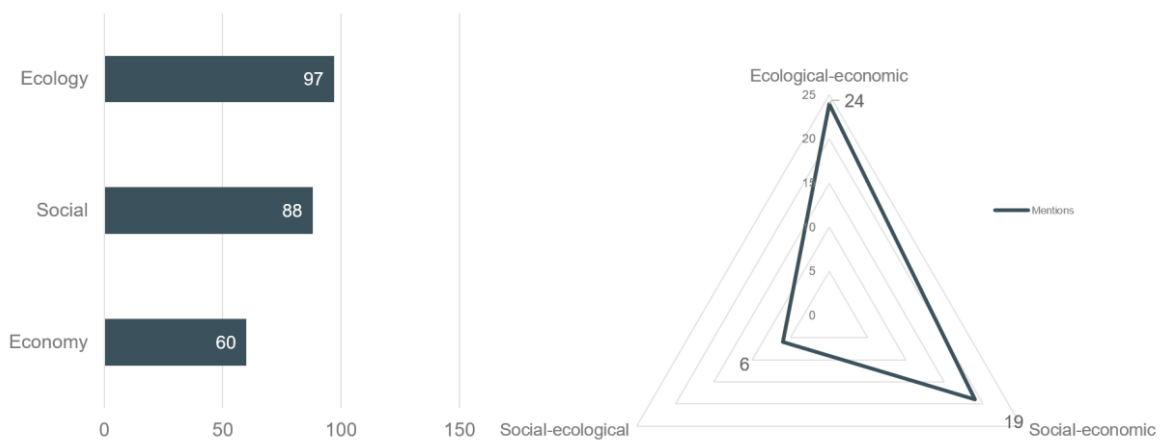

**Fig. 3** Dimensions of the triple bottom line mentioned (left) and arising conflicts (right)[1]

*Relevance of sustainability in departments, processes, and strategies*

---

[1] The figures in this publication show the total number of codings, as sometimes, multiple codings were awarded within the same publication when mentioned in a different context.

In the material, the most frequently discussed departments that were responsible to integrate sustainability into practice were sales and marketing, finance, and the sustainability department (Fig. 4). Less frequently, procurement and HR were mentioned.

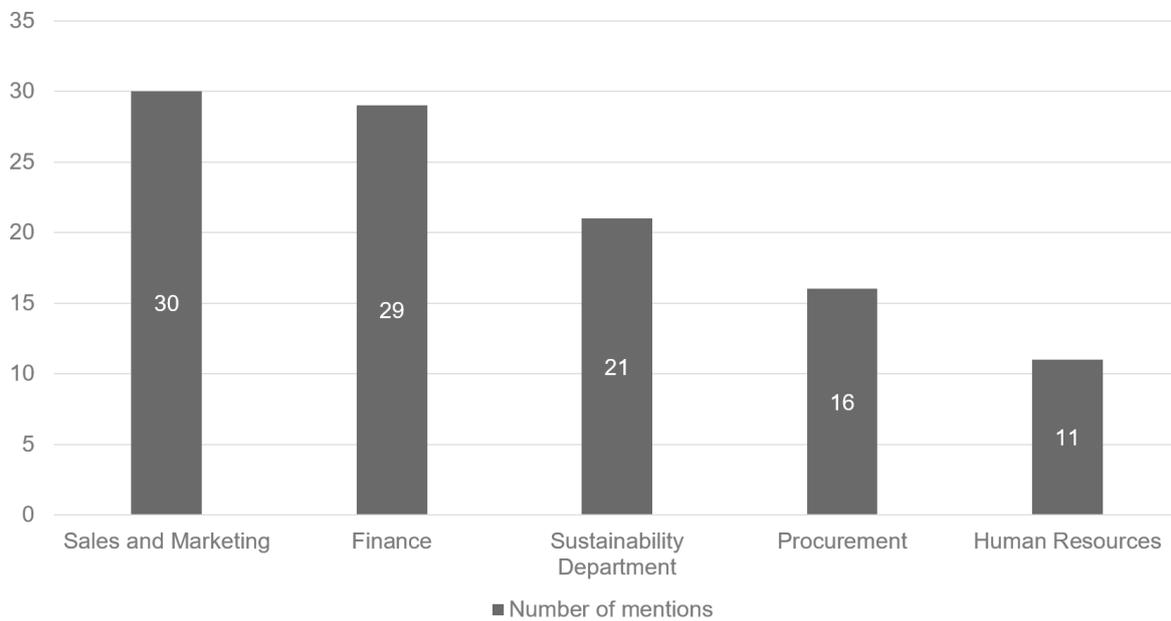

**Fig. 4** Integration of sustainability on a departmental level (frequency of mentions)

The sustainability department was mentioned as a possible point of interaction with controlling/accounting because they aggregate corporate information from different areas of the business, generating an overview of environmental, social and economic sustainability progress (Isensee & Michel, 2011). However, the degree of cooperation between the interfaces varies. While not all businesses have a sustainability department, those that do usually are the main locus of sustainability information for all other business areas (Michel et al., 2014). In contrast, employees in controlling and accounting do not see themselves as relevant stakeholders of the sustainability department and rarely keep up with their developments, unless they have a personal interest in the topic (Kämmler-Burrak & Bauer, 2022; Stehle and Stelkens, 2018).

In procurement, the sustainability efforts that are connected to controlling and accounting usually focus on the corporate supply chain's emissions and resource consumption (Ferreira et al., 2010; Glanze, 2022; Krause, 2016). However, some publications mention progress in the sense of design for sustainability, procurement for sustainability or the development of more efficient and environmentally friendly means of production (Greiling & Ther, 2010; Isensee & Michel, 2011; Schaltegger, 2020; Schaltegger & Zvezdov, 2011).

Sustainability in HR mostly focuses on the social dimension through employee development and training, as well as the controlling of employee KPIs (Greiling & Ther, 2010; Weber et al., 2010). They are a stakeholder of the sustainability department because for an increasing number of potential employees, a company's sustainability efforts are relevant in the application process. HR needs sustainability information to market the roles to applicants (Braun, 2019). It is also in the interest of the company that employees act in line with the corporate strategy. This also includes sustainability-specific aspects of the strategy. It is therefore necessary for HR to also offer further training in this area and to do so in cooperation with the sustainability department (Lacy et al., 2009). Regarding controlling, HR also records data relevant to the social dimension of sustainability, such as absence and sickness rates, workplace incidents and personal development hours, amongst others (Schwarzmaier, 2015).

Avoiding green washing and the development of an 'honest' marketing mix dominate the codings in the field of sales and marketing (Gebauer, 2013; Schaltegger & Zvezdov, 2011; Schwarzmaier, 2013). Braun (2019) even calls some corporate sustainability (CSR) reports a mere marketing tool that is used strategically as a means of corporate communications. These reports provide little value to stakeholders as they lack substance (Gaggl, 2021). To enforce truly sustainable marketing efforts, it should be seen as an opportunity to realize the corporate sustainability strategy, i.e. by communicating on internal sustainability efforts (Schwarzmaier, 2013; Weigand, 2020).

In the finance department, most codings relate to the area of controlling (Fig. 5). These were coded when the area was mentioned in a sustainability context. The other areas like external and internal accounting (the latter represented by the different areas of cost accounting and investment accounting) were mentioned far less frequently, although explicitly included in the search terms (Table 1).

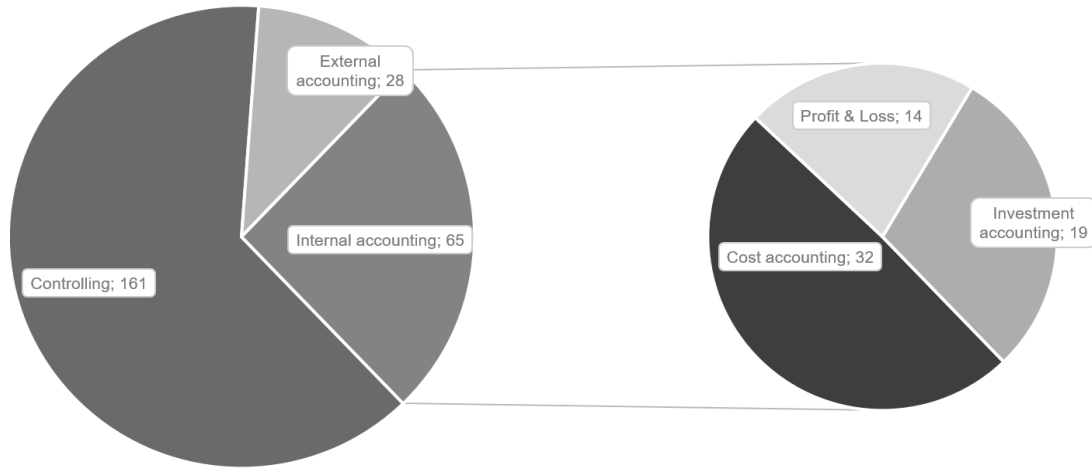

**Fig. 5** Coding distribution across different accounting and controlling departments with an overview on the left and a zoom into internal accounting on the right

Most texts did not discuss new sustainability-oriented instruments for controlling but rather applied traditional tools to this new context (Wellbrock et al., 2020). This stems from the idea that while the main focus remains to achieve economical sustainability or persistence, rather, social, and ecological goals are becoming more relevant. Some traditional tools that are mentioned are the ecology-oriented SWOT analysis, portfolio analysis, ABC-scoring, eco-efficiency indicators or ecological life cycle assessments (Greiling & Ther, 2010). Some publications specify controlling tools that are unique to the issue of sustainability, for example the 'sustainable value' or the 'sustainability balanced scorecard' (Figge & Hahn, 2004; Gaggl, 2021; Heimel & Momberg, 2021; Hubbard, 2009; Wellbrock et al., 2020). Fischer et al. (2010) and Michel et al. (2014) highlight the need of new KPIs that include the triple bottom line approach. Thus, new KPI systems need to be developed to ensure number-based reporting (Lühn et al., 2022). According to Aras and Crowther (2009) and Rönz and Ryba (2012) however, companies need guidance to develop these KPIs, as many sustainability aspects are difficult to quantify/put into a financial context.

Some publications go further and discuss the connection between controlling and CSR departments, as both collect relevant data that requires profound knowledge about sustainability topics (Rönz & Ryba, 2012). Since most controlling employees do not possess extensive knowledge about sustainability topics besides what is required for their specific area of control, a cooperation is needed to successfully develop and implement a corporate sustainability strategy (Endenich & Trapp, 2019). These aspects uncover an issue that is frequent with this type of literature review – it would be useful to include best practice examples for the demands that the publications put onto industry practice as the issues they discuss and the solutions they suggest remain quite complex.

In both external and internal accounting, the sustainability aspects were mostly discussed as supplemental to the regulations companies already have to abide by (Günther & Stechemesser, 2011; Sulaiman & Mokhtar, 2012). In external accounting, most codings related to hardships that companies who make more sustainable choices face when it comes to financial accounting and reporting. Sulaiman and Mokhtar (2012) specify that professional accounting bodies must require monetary information to be reported in annual sustainability reports, because only this will motivate companies to be serious about the integration of environmental issues into their accounting systems. Schwarzmaier (2015) adds that companies that prepare reports in Germany can use accounting instruments such as social balance sheets in addition to the traditional management reports, although they remain voluntary. Regarding internal accounting, a broader scope for action can be identified (Fig. 5, right side). In cost accounting, it is an option to internalize costs that before have been externalized, e.g., $CO_2$ compensation. Some companies go a step further and prepare an ecology-oriented profit and loss statement that internalizes effects on ecology and values them at an internal cost rate (Michel et al., 2014).

*Controlling instruments and their sustainable alternatives*

The tools that are mentioned in controlling and accounting codings extend from more general calculations like KPIs with (n=19) and without a sustainability focus (SKPI, n=16) to more specific ones, such as eco efficiency analysis (n=3), social return on investment (SROI, n=2), sustainability balanced scorecard (SBSC, n=19) and sustainable value (n=9). The SBSC approach was the most frequently mentioned sustainable tool with n=19 mentions. It was usually discussed as a means to operationalize sustainability goals for companies and departments, thus more useful as a starting point than from an advanced position.

Interestingly, the general KPIs were rather specific control systems like economic value added (EVA) and total ROI, while with SKPIs, older publications stressed that companies need to develop their own KPIs or adapt existing ones, such as the ecology value added (Lux

& Olbert-Bock, 2016; Schulze & Thomas, 2012; Schwarzmaier, 2013). Newer publications suggested SKPIs such as the recording of CO2 emissions, circular transition indicators or sustainable value added (Gaggl, 2021; Kämmler-Burrak & Bauer, 2022; Lühn et al., 2022). Additionally, newer publications also stressed the issue of lacking data availability and standardization issues (Kämmler-Burrak & Bauer, 2022). In recent years, there has been an influx of KPIs and tools in sustainability controlling and accounting, most of which companies have developed themselves by benchmarking (Heimel & Momberg, 2021; Rönz & Ryba, 2012).

*Sustainability within the supply chain*

In addition to sustainability efforts on a departmental level, some aspects affect some or all parts of a business's supply chain. A general distinction is made between management/control processes (where decisions are made), key processes (where value is created) and support processes (where value creation and decision-making are enabled) (Krajewski, 2021). Most frequently, management and control processes (n=94) as well as the entire supply chain (n=53) were coded. Key and support processes were not coded as often (n=29 and n=14, respectively).

Publications that discuss sustainability in management processes suggest that the topic can be integrated in multiple ways (Lacy et al., 2009; Olbert-Bock et al., 2015; Stehle & Stelkens, 2018). Sustainability should be integrated both top-down and bottom-up by engaging in continuous stakeholder dialogue. However, a significant amount of data is needed for adequate sustainability performance management within these processes. Another option is to incentivize sustainable decision-making (Schäffer, 2022). Additionally, non-financial information should be as relevant in management decisions as financial information and thus discussed as such in corresponding contexts (Sulaiman & Mokhtar, 2012). Some publications also discuss key processes such as production, logistics and procurement as appropriate options to promote sustainability (Ferreira et al., 2010; Fischer et al., 2010; Weber et al., 2010). This could be the reduction of scrap products and materials, resource, and energy consumption, as well as the extension of product life. In support processes, sustainability takes on the form of employee habit change and the sustainability of the company itself (Krause, 2016; Weigand, 2020). This extends to aspects such as using recycled printing paper (or not printing at all), being conscious of cloud data and server energy usage, recycling inhouse, using regenerative energy to power production sites and offices and donating to charity. Compared to a sustainable orientation of key processes and the entire supply chain, these changes usually only have a small impact on the overall sustainability of a company. Publications discussing a company's total supply chain mostly focus on the social dimension of sustainability by describing the impact on local labor laws and even environmental protection laws at production sites (Gaggl, 2021; Hubbard, 2009; Klute-Wenig & Refflinghaus, 2015; Rodriguez-Olalla & Aviles-Palacios, 2017; Schulz, 2014).

*Reasons to move towards sustainability*

Regarding the reasoning behind the implementation of sustainability, most publications mentioned either regulatory or profit-oriented reasons (39 and 40 codings, respectively). Regulatory reasons often touched on local laws regarding non-financial/integrated reporting and CSR or labor laws (Horváth et al., 2017; Klute-Wenig & Refflinghaus, 2015; Nikolic et al., 2020; Pandit & Rubenfield, 2016; Schwarzmaier, 2013). Enforcing more sustainable business practices for social benefits or out of individual conviction/philanthropy were coded less frequently (16 and 24 mentions, respectively), some touch on small and medium-sized enterprises (SMEs) as more prone for a holistic sustainable transformation (Aras & Crowther, 2009; Kaldschmidt, 2011; Schmidpeter & Günther, 2013; Weber et al., 2010).

Additionally, most texts discuss a strategic integration of sustainability (n=81 codings). Operative and tactical periods are mentioned less often (Fig. 6). This is an interesting finding when relating back to Fig. 4, where many texts touched on marketing as the main domain for sustainability efforts.

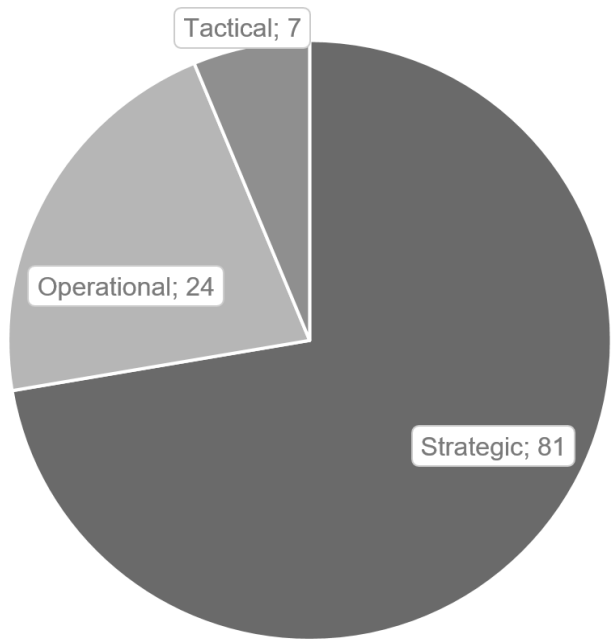

**Fig. 6** Mentioned period for sustainable transformation

*Sustainability reporting items as well as opportunities and risks of a sustainable transformation*

Regarding the non-financial data that companies reported on, a few trends can be identified in the publications (Fig. 7). Most publications mention that companies use the Global Reporting Initiative (GRI) standards and focus on emissions, the exploitation of resources and work conditions (Gawenko et al., 2020; Pandit & Rubenfield, 2016; Rönz & Ryba, 2012; Weber et al., 2010). Less emphasis is put on creating equal opportunities, keeping human rights or fighting lobbyism and corruption (Schier & Bonnländer, 2018; Schmitz, 2021).

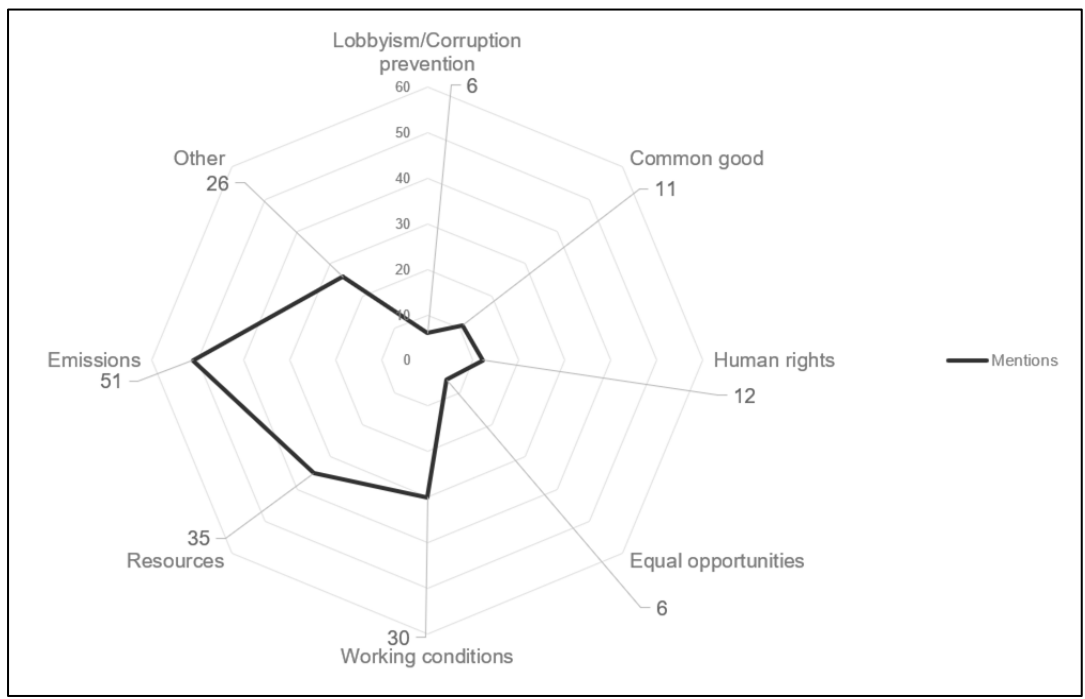

**Fig. 7** CSR reporting items mentioned by publications

This might be related to the reasons why companies integrate sustainability in the first place, since the trending reporting items are not as heavily regulated as the other items. Thus, they need to be reported on separately. Since most of the publications are from Germany and labor laws are on the stricter side here, these aspects are not of as much interest to companies as their emissions, the use of resources or working conditions. However, it can be argued that the less frequently mentioned aspects should be considered when looking at the entire supply chain (which, arguably, is necessary when striving for a truly sustainable company), which was not a central focus in most of the publications.

Publications that encourage a sustainable transformation through accounting and controlling commonly identified opportunities and challenges regarding the process (Fig. 8). Cost reductions and competitive advantages were the opportunities named most frequently, this again matches with the motivating factors and conflicts previously identified (Ferreira et al., 2010; Klute-Wenig & Refflinghaus, 2015; Schulz, 2014; Weber et al., 2010). The aspects of employee retention and contribution to combating climate change were mentioned least frequently (Braun, 2019). Regarding the challenges, by far the most often mentioned aspect was complexity (Gaggl, 2021; Greiling & Ther, 2010; Hubbard, 2009; Schäffer, 2022). This makes sense when looking at the fast-moving, detailed supply chains that companies nowadays have. In addition, rules and regulations differ across the planet and, most importantly, there is not always a clearly 'most sustainable option,' as sustainability as a lot of variables and companies face insecurities when pursuing a sustainable transformation. Interestingly, costs/financial investments were not named as major challenges in the publications (Preiß, 2019; Sulaiman & Mokhtar, 2012).

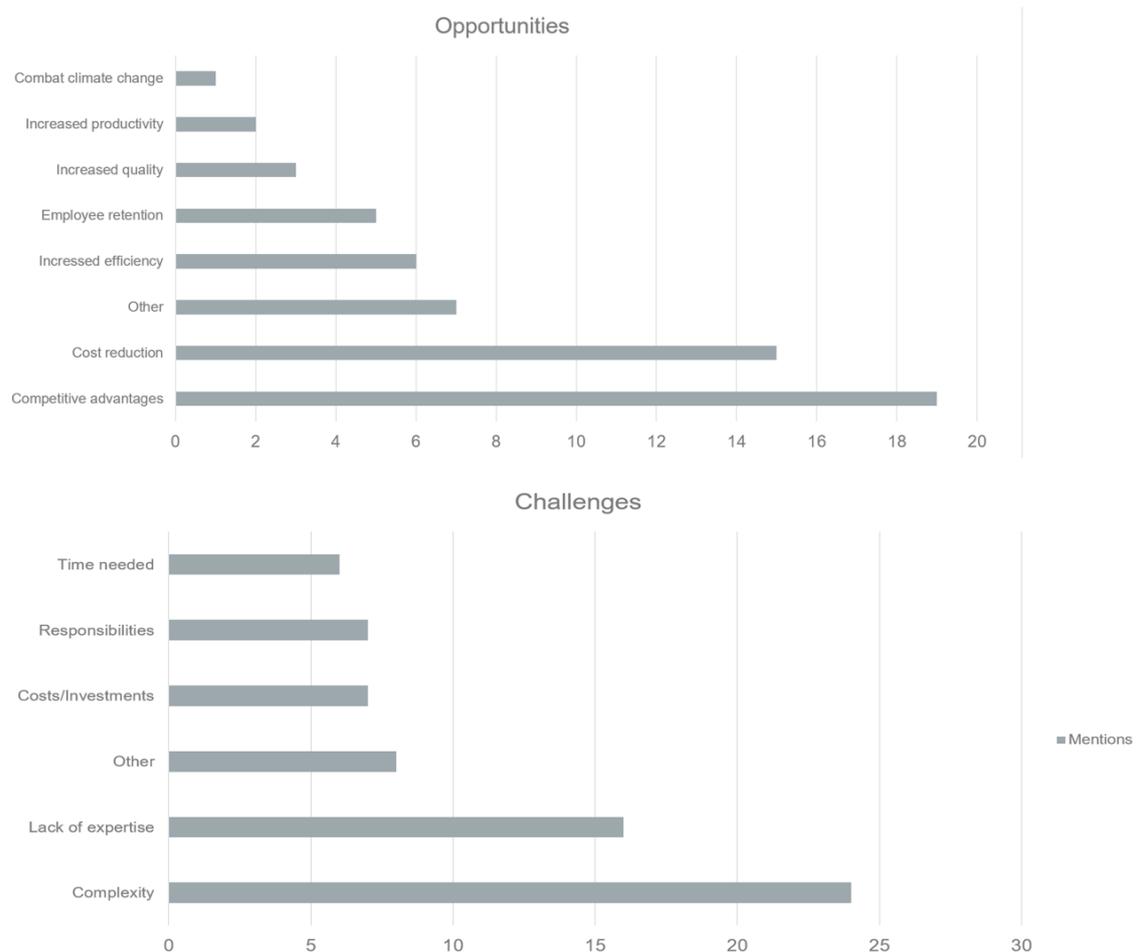

**Fig. 8** Opportunities (top) and challenges (bottom) regarding a sustainable transformation

*RQ2: Which sustainability-focused competencies are discussed in accounting and controlling research and practice?*

Employees in controlling and accounting need to be well educated to face the changing conditions in the corporate environment. These skills can be general or with a focus on sustainability, since some skills are easily adaptable for different contexts (e.g., researching alternatives for products or developing new strategies), whereas others are only needed in specific sustainability-focused situations (e.g., the calculation of $CO_2$ emissions or the corporate carbon footprint). In the n=79 publications, professional and methodological skills were discussed the most (n=114), personal skills 23 times and social skills the least (n=12). Not all of these were made in a sustainability-oriented context although made in a publication aiming at sustainability in accounting and/or controlling.

To further illustrate this difference, regarding professional and methodological skills, 83% of the mentions were made with regard to sustainability, this applies to only 56% of personal skills and 66% of social skills. Apparently, the literature deems professional and methodological skills as more necessary when discussing a sustainable transformation (Hoekstra, 2020; Kämmler-Burrak & Bauer, 2022; Sneidere & Bumane, 2019; Tsiligiris & Bowyer, 2021). In addition, the fact that not all skills are related to sustainability supports the idea that there are some aspects of vocational action competence that can already be adapted to reflect more sustainable choices, without explicitly including the topic (Fischer et al., 2010; Mödritscher & Wall, 2021; Schubert & Gerhardt, 2021; Sneidere & Bumane, 2019).

*Professional and methodological competence*

Results from the dimension of professional and methodological skills can be clustered into three key aspects: 1) sustainability knowledge 2) knowledge about tools and KPIs and 3) data and interface management.

The first aspect touches on the development of professional knowledge about sustainability. Employees in controlling and accounting need to identify intersections between their area of expertise and potential sustainability aspects to analyze information potentially relevant for sustainability controlling processes (Gaggl, 2021). It is not required for them to become sustainability experts, but there is nuance: employees in controlling and accounting should be open to include sustainability-oriented tools and processes into their repertoire and be available to take on learning opportunities (Lühn et al., 2022; Schulz, 2014; Stehle & Stelkens, 2018). Especially the methodological skill can be developed to include the transformation of ecological and social effects into financial units (Weber et al., 2010). To aid in this, companies need to train controllers as additional 'sustainability managers' by developing their general skill of preparing management information (Schulz, 2014; Schulze & Thomas, 2012). Lastly, these employees need to be able to communicate the performance of management accounting reports or other regular external reporting and realize their value for controlling processes (Weber et al., 2010).

Regarding the second aspect, publications mention three relevant perspectives. Those concepts that are currently used need to be reevaluated for their suitability regarding a sustainable transformation. Thus, employees need to be able to measure, analyze, report and evaluate financial data to quantify non-financial information, i.e. by adapting existing KPIs and/or tools (Clarke & O'Neill, 2005). In doing so, they need to apply legal frameworks and keep up with industry standards and best practices. When adaptation is not an option, new concepts need to be developed by defining new relevant KPIs that gather the corporate environmental and social impacts (Stehle & Stelkens, 2018). Necessary skills include benchmarking, risk analysis, stakeholder surveys and external ratings (Kämmler-Burrak & Bauer, 2022). When finally, all tools are up to date, they need to be competent in the use of the new tools, thus broadening their skills in measuring sustainability-relevant information (Hartmann et al., 2016). Then, employees should consider the fit with the corporate sustainability strategy and control accordingly. This might require new workflows (Schulze & Thomas, 2012).

Data and interface management considers the support of other potentially relevant departments and processes. Not all information relevant to sustainability is aggregated in a single department. Thus, departmental cooperation is required. Traditionally, employees in controlling and accounting do not consider sustainability to be central to financial processes. When striving for sustainable transformation, this value needs to shift. Some publications suggest collaboration between the sustainability experts and those employees who now need to include sustainability in their activities (Fischer et al., 2010; Isensee & Michel, 2011). This can be done through an additional or integrated sustainability department. This reduces specialized technical knowledge that employees in controlling and accounting – who are already highly specialized – need to develop (Schaltegger, 2020). Overall, employees in controlling and accounting will take on an active supporting role in sustainable transformations, as they possess all data that is needed to make holistically sustainable decisions (Weber & Schäffer, 2016).

*Social and personal competence*

Regarding social skills, publications primarily focus on cooperation, collaboration, and communication. It is necessary for employees to use high-quality data as a basis and keep it up to date by collaborating with other departments through empathy and taking initiative (Gebauer, 2013; Schubert & Gerhardt, 2021). This extends to the collection of relevant data from other departments, including the sustainability department (Fischer et al., 2010). The personal skills that publications discuss can be divided into ethical guidelines and respecting values as well as identifying complexities and being willing to learn. Regardless of their personal opinion on sustainability, employees need to respect company values and guidelines in their recommendations (Walinska & Dobroszek, 2021) and produce comparable, reliable and consistent data and reports (Arora et al., 2022). Additionally, they need to put problems into larger contexts and derive implications for their actions (D'Amato et al., 2009).

The competence requirements clearly show that controllers need to expand their existing skills: a certain level of sustainability knowledge is required as well as the ability to navigate sustainability-oriented tools and key figures by adapting them or developing new ones. Data management is also important, as non-financial data may need to be transformed or communicated. Since sustainability data occurs everywhere in the company, controllers also have to manage interfaces. Overall, controllers do not need to be experts, but they do need to cooperate.

**Discussion**

*Sustainability in the Occupational Reality of Employees in Controlling and Accounting*

The exploration of sustainability in accounting and controlling literature reveals a dynamic shift from a traditional perspective, which primarily aimed at securing the long-term existence of companies, to an integrated approach aligned with the three-pillar model. Notably, sustainability discussions have expanded beyond separate sustainability management and controlling departments, indicating a need for a more integrative approach within organizations. The literature highlights the challenges faced by companies, including a lack of expertise, financial resources, and industry interest, emphasizing the multifaceted nature of sustainability integration (Gaggl, 2021; Nikolic et al., 2020; Okwuosa, 2020). Relevant tools and processes in controlling predominantly focus on KPIs, cost management, and sustainability KPIs, underscoring the need for a comprehensive understanding of both financial and non-financial aspects. Furthermore, the results suggest the importance of sustainability-oriented information and tools that integrate sustainability into mainstream controlling processes rather than segregating it into a distinct sustainability controlling function. As a result, sustainability seems to not have arrived on the level of the individual employee but remains an issue for management. For example: How is sales personnel supposed to consult potential customers on the sustainability of their products when they themselves have no real concept of the sustainability of their products? For an educated consultation, data and product information are needed from management.

*Organizational Levels and Motivations*

The identified integration of sustainability initiatives spans across various organizational levels, ranging from dedicated sustainability departments to functions within sales, marketing, PR, and finance, here most often it is sustainability controlling. The predominant motivations driving this integration are legal compliance and profit optimization. The strategic implementation of sustainability measures is dominated by the economic dimension, emphasizing the prevalence of economic goal conflicts. The ecological dimension is affected through the reduction of emissions and resources, while the social dimension mostly considers working conditions, often interconnected with legal obligations such as labor laws.

*Operational Activities in Controlling and Accounting*

Controlling activities predominantly concentrate on KPI with and without sustainability focus and their connection to cost management. Noteworthy is the emphasis on capturing sustainability-relevant management information across the company, utilizing tools that consider sustainability holistically, instead of relying on separate sustainability controlling entities. In the domain of accounting, challenges include the incorporation of non-monetary information, such as the accrual of financial information, and the external reporting of non-financial aspects. Proposed solutions include amortization through funds, the utilization of social balance sheets, or environmental investment calculations as complementary tools alongside traditional instruments like the income statement.

*Necessary Competencies for Vocational Education*

The discussion on sustainability-focused competencies in accounting and controlling research and practice underscores the critical role of change agents in driving corporate transformation towards sustainability. These change agents, often termed as "controllers," require competencies such as sustainability-related learning readiness, collaboration, identification, and adaptation of relevant tools and KPIs, understanding and addressing complexities and conflicts, and weighing ethical principles. This suggests that even in organizations with dedicated sustainability departments, controllers need to possess sustainability-related competencies, indicating the pervasive nature of sustainability across different facets of business operations (Pandit & Rubenfield 2016; Walinska & Dobroszek, 2021).

The identified competencies extend beyond technical proficiency to include sustainability-related learning readiness, collaboration, and cooperation. Professionals need to identify and potentially develop or adapt relevant tools and metrics, recognizing the evolving nature of sustainability metrics. The literature focuses on the importance of understanding complexities and conflicts to be enable students to propose viable solutions as change agents (Isensee & Michel, 2011; Sneidere & Bumane, 2019; Walinska & Dobroszek, 2021; Weber & Schäffer, 2016). In addition, personal competence should balance personal and organizational (ethical) principles, which are both crucial for navigating sustainability decision-making (Schubert & Gerhardt, 2021; Walinska & Dobroszek, 2021).

Synthesizing these insights, the results point towards a paradigm shift in organizational practices, where sustainability is a strategic imperative while considering economic implications. The competencies highlighted serve as a starting point to develop critical implications for vocational education. These indicate the need for a comprehensive and dynamic curriculum that not only imparts technical knowledge but cultivates a holistic perspective on the complexities and ethical dimensions of sustainable decision-making as change agents.

*Implications for accounting and controlling vocational education*

Combining the results described above and the central arguments from the discussion with the critical perspective on sustainability in accounting and controlling developed in the introduction, the following implications should be considered:

- The integration of sustainability in **accounting and controlling learning scenarios** to connect the increasing relevance that sustainability has in decision-making processes. Sustainability is reaching controlling departments; thus, classes should include the concept across study programs. It is no longer enough to only create connection whenever it 'fits' with the school-wide curriculum – rather, teachers need to view sustainability in controlling as an intersectional topic that affects the topic in multiple dimensions and discuss potential conflicts.

- Connecting to this, teachers need to be aware of **alternative controlling and accounting tools and KPIs** that they can use as a substitute or add to their repertoire. They need to be able to research relevant additions to the school-wide curriculum and integrate them into learning scenarios, i.e., the sustainability balanced scorecard approach or the application of the sustainable value metric as a substitute or an addition to the ROI. In this context, controlling classes should also consider the importance of non-financial KPIs and their potential for learning opportunities as they may aid to illustrate conflicts and foster valuable learning opportunities as they connect the dimensions of economics, business management and legal. Since students might have to adapt tools and KPIs to integrate sustainability in the future, including this whenever possible seems beneficial, too. This aspect extends to teacher education, where sustainability currently only has minor impact (Evans, 2019; Trampe, 2023). Relevant specific and general perspectives on the respective subjects should be developed and integrated into educational programs.

- Curricula should include the issue of **economic sustainability as the primary goal** to use sustainability conflicts in a didactically beneficial manner, rather than to make the students feel demotivated. Students will need to develop a set of competences that helps them drive for organizational sustainable change as part of their vocational action competence. For this, it might be helpful to discuss alternatives and best practice examples.

- A deepened **business process orientation** that focuses on management and key processes to enable students to take on a multidimensional approach. The aim is to bridge the gap between the analysis of situations and the deduction of didactical suggestions and necessary changes. German vocational business education is, in theory, inherently process-oriented; thus, sustainability integrated processes may uncover potential starting points for lesson planning. When discussing the entire supply chain, sustainability-oriented aspects will have to be considered to depict the corporate situation holistically. In addition, discussing business processes helps illustrate conflicts between sustainability dimensions because it illustrates cause-and-effect chains that sometimes are too abstract for students who are used to simplified learning situations in class.

- **Ethical aspects** need to be discussed to encourage students to push for change as sustainable change agents. In some companies, students experience a dissonance between their subjective ethical values and corporate practice. They then need to make decisions that are beneficial for the company, in the process they must compromise their own values. This becomes particularly clear when looking at the motivation to push for sustainability, where profit and legal reasons were the main ones. With more students becoming conscious about sustainability and ethics, the disparity might cause issues that need to be solved, for example by teachers that openly discuss potential conflicts and encourage students to think of solutions in a safe space, i.e., the classroom. The guiding principle of a "green controlling employee" (see i.e., Stehle & Stelkens, 2018) might be helpful, as it includes the modern perspective of the "thinking accountant/controller" on accounting and controlling, instead of the traditional "practicing accountant/controller" (Reinisch, 1996).

*Limitations*

There are some limitations to this review that need to be discussed. Although with PRISMA, we used a standardized strategy, we cannot control 'intelligent' search algorithms that use geolocalization (particularly with Google Scholar) and smart indexing. This might explain the uneven distribution of the two languages within the sample.

In addition to this rather technical aspect, there is a content-related limitation as in German concepts, controlling only comprises internal accounting and is decision-oriented, whereas internationally, a more integrated approach is common (Beckers, 2010). Thus, the data basis controllers use is different. These variations have grown historically and hence contributed to different views on the roles of controllers.

Lastly, the share of publications discussing SME is small, which limits the significance of the results for students who work in this context. Here in Germany, almost 70% of students do their apprenticeship in SME (Bundesinstitut für Berufsbildung, 2023, 196 f.), thus, their workplaces might not yet integrate sustainability into controlling and/or accounting (Alsdorf et al., 2023; Martins et al., 2022). However, apprenticeships equip students with a solid base for future workplace changes through the development of vocational action competence. It is likely that most students who are currently doing an apprenticeship will have to face sustainability-focused aspects in these areas at some point during their career.

**Conclusion**

This research paper sheds light on the role of sustainability in accounting and controlling, with a particular emphasis on its implications for vocational education. The study provides valuable insights into how sustainability is discussed in accounting and controlling literature and the competencies essential for practitioners in this field:

In the context of vocational education, the study proposes implications for accounting and controlling teachers. It advocates for the integration of sustainability into learning scenarios to align with the increasing relevance of sustainability in decision-making processes. The paper emphasizes the need for teachers to be aware of alternative tools and KPIs, fostering a proactive approach to researching and integrating relevant additions into the curriculum. Additionally, ethical considerations are highlighted, acknowledging the potential dissonance between personal values and corporate practices that students may encounter in their future careers. The study encourages a deepened business process orientation to enable students to take on a multidimensional approach and emphasizes the development of a set of competences that prepares students to function as sustainable change agents.

The research not only contributes to understanding where companies are currently at regarding sustainability in accounting and controlling literature but also provides practical guidance for vocational education, urging educators to adapt their approaches to equip students with the necessary competencies for a future where sustainability is integral to corporate decision-making.

The implications drawn from this research paper hold relevance within the context of the vocational education system in Germany. German vocational business education has long been renowned for its emphasis on practical training and its integration into the workforce. As the study highlights the increasing importance of sustainability in accounting and controlling, it becomes imperative to consider how these findings can be effectively incorporated into the German vocational education curriculum. Given that almost 70% of students in Germany undergo their apprenticeships in small and medium-sized enterprises (SMEs), the paper acknowledges the potential disconnect between the theoretical integration of sustainability in education and its practical application in workplaces that may not yet fully embrace sustainability in controlling and accounting practices. Despite this challenge, the study highlights the significance of vocational education in equipping students with a solid foundation for future workplace changes, emphasizing the development of vocational action competence. As sustainability becomes a central theme in corporate decision-making, the German vocational education system must adapt its curriculum to ensure that students are not only well-versed in traditional accounting and controlling methods but also possess the competencies required to navigate and contribute to sustainable business practices in real-world scenarios. To do so, teachers need to be adequately trained, too.

Winther, E. & Achtenhagen, F., (2008). Kompetenzstrukturmodell für die kaufmännische Bildung. *Zeitschrift für Berufs- und Wirtschaftspädagogik, 104*(4), 511–538. https://doi.org/10.25162/zbw-2008-0028

Wöhe, G., Döring, U. & Brösel, G. (2020). *Einführung in die allgemeine Betriebswirtschaftslehre*. Verlag Franz Vahlen.

World Commission on Environment and Development (1987). *Report of the world commission on environment and development: Our common future.* http://www.un-documents.net/wced-ocf.htm. Accessed 05 April 2024.
36

# Appendix A. Analytical Framework with Sub-Categories

**Table A.1**

| Main category | Level 1 sub-category | Level 2 sub-category |
|---|---|---|
| Competence requirements | Professional and methodological competence | |
| | Social competence | |
| | Self-competence | |
| | Focus on sustainability | |
| Activities | Controlling | |
| | Internal accounting | |
| | Integrated reporting | |
| | External accounting | |
| Approach to sustainability | Level of business areas | Procurement, HR, PR, marketing, finance |
| | Level of business processes | Control, core, support processes, entire supply chain |
| | Motivation to approach topic | Legislation, profit optimization, social recognition, philanthropy |
| | Period of integration | Strategic (long-term), operational (medium-term), tactical (short-term) |
| | Sustainability dimensions | Ecology, Economy, Social |
| | Sustainability strategies | Efficiency, consistency, sufficiency |
| | Sustainability conflicts | Ecology-social, economy-social, economy-ecology |
| Relevant certificates | Certificates | SA8000, ISO5000, ISO26000, EMASII, ISO9000, others |
| and frameworks, tools, and methods | Reporting standards (Contents, not frameworks) | Corruption prevention, human rights, public welfare, equal opportunities, working conditions, resources, emissions |
| | Tools, Concepts, KPIs | Sustainable Value, Sustainability Balanced Scorecard, Social Return on Investment, eco-efficiency analysis, sustainability key performance indicators |

**Appendix B. List of coded publications**

| Author | Year | Title | Journal/Publisher |
|---|---|---|---|
| **Altenburger, Reinhard** | 2022 | Strategisches CSR-Controlling in Familienunternehmen | Springer Nature |
| Anderson, Kai; Sommer, Carina; Fassino, Gina | 2022 | Kann HR Nachhaltigkeit? | Personalmagazin |
| Aras, Güler; Crowther, David | 2009 | Corporate Sustainability Reporting: A Study in Disingenuity? | Journal of Business Ethics |
| Arora, Mitali Panchal; Lodhia, Sumit; Stone, Gerard | 2022 | Enablers and barriers to the involvement of accountants in integrated reporting | Mediterranean Accountancy Research |
| Asenkerschbaumer, Stefan; Watterott, Richard | 2011 | Controlling-Steuerungsimpulse für ein Mehr an Nachhaltigkeit am Beispiel Bosch | Controlling |
| Bernatzky, Simone; Endenich, Christoph; Wömpener, Andreas | 2018 | Zur Integration von Nachhaltigkeit in das Controlling - Eine empirische Analyse | Betriebswirtschaftliche Forschung und Praxis |
| Biel, Alfred | 2020 | Nachhaltigkeit ein Controlling-Thema? | CONTROLLER Magazin |

| Authors | Year | Title | Journal |
| --- | --- | --- | --- |
| Biswas, Sharlene; O'Grady, Winnie | 2016 | Using external environmental reporting to embed sustainability into organisational practices | Accounting Research Journal |
| Bogicevic, Jasmina; Domanovic, Violeta; Krstic, Bojan | 2016 | The role of financial and non-financial performance indicators in enterprise sustainability evaluation | Ekonomika |
| Braun, Sabine | 2019 | Nachhaltigkeitsberichterstattung: Im Spannungsfeld unterschiedlicher Anforderungen und Interessen | Ökologisches Wirtschaften |
| Burritt, Roger; Schaltegger, Stefan | 2001 | Eco-efficiency in corporate budgeting | Environmental Management and Health |
| Clarke, Kevin; O'Neill, Sharron | 2005 | Is the Environmental Professional … an Accountant? | Greener Management International |
| D'Amato, Alessia; Roome, Nigel; Lenssen, G. | 2009 | Toward an integrated model of leadership for corporate responsibility and sustainable development: a process model of corporate responsibility beyond management innovation | Corporate Governance: The International Journal of Effective Board Performance |
| Dumitru, Mădălina; JINGA, Gabriel | 2015 | Integrated Reporting Practice for Sustainable Business: A Case Study | Audit Financiar |
| Endenich, Christoph; Trapp, Rouven | 2022 | Nachhaltigkeitscontrolling in Klein- und Mittelunternehmen | Springer Nature |

| Faupel, Christian; Stremmel, Florian | 2011 | Berücksichtigung von Nachhaltigkeit im Rahmen einer wertorientierten Unternehmensführung | Controlling & Management |
| --- | --- | --- | --- |
| Ferreira, Aldónio; Moulang, Carly; Hendro, Bayu | 2010 | Environmental management accounting and innovation: an exploratory analysis | Accounting, Auditing & Accountability Journal |
| Figge, Frank; Hahn, Tobias | 2004 | Sustainable Value Added—measuring corporate contributions to sustainability beyond eco-efficiency | Ecological Economics |
| Figge, Frank; Hahn, Tobias; Schaltegger, Stefan; Wagner, Marcus | 2002 | The Sustainability Balanced Scorecard - linking sustainability management to business strategy | Business Strategy and the Environment |
| Fischer, Stephan; Schmitz, Anja; Knepel, Kirke | 2014 | Auswahl von nachhaltig handelnden Mitarbeitern - vom Kompetenzmodell zur Auswahlmethode | Wirtschaftspsychologie |
| Fischer, Thomas M.; Huber, Robert; Sawczyn, Angelika | 2010 | Nachhaltige Unternehmensführung als Herausforderung für das Controlling. | Controlling |
| Gaggl, Philipp | 2021 | Nachhaltigkeitscontrolling: Wie nichtfinanzielle Informationen zum Werttreiber werden | Springer Gabler |

| Autor(en) | Jahr | Titel | Quelle |
|---|---|---|---|
| Gänßlen, Siegfried; Kraus, Udo; Ette, Daniel | 2011 | Green Controlling - Green Profit - Nachhaltigkeitscontrolling bei Hansgrohe | Controlling |
| Gawenko, Wladislav; Richter, Fanny; Hinz, Michael; Götze, Uwe | 2020 | Interne Ansätze zur Nachhaltigkeitsbewertung in der externen Berichterstattung. Konzeptionelle und empirische Analyse der DAX-Unternehmen | Die Unternehmung |
| Gebauer, Jana | 2013 | Die Zukunft der Nachhaltigkeitsberichterstattung III und IV/IV: Glaubwürdigkeit können nur die Unternehmen selbst herstellen | Ökologisches Wirtschaften |
| Glanze, Eva | 2022 | Kompass für Green Change | OrganisationsEntwicklung |
| Greiling, Dorothea; Slacik, Johannes | 2022 | Nachhaltigkeitsberichterstattung von Elektrizitätsversorgungsunternehmen: Ein internationaler Vergleich | Springer Nature |
| Greiling, Dorothea; Ther, Daniela | 2010 | Leistungsfähigkeit des Sustainable Value-Ansatzes als Instrument des Sustainability Controlling | Springer |
| Günther, Edeltraud; Stechemesser, Kristin | 2011 | Instrumente des Green Controllings: ein Blick zurück, ein Blick nach vorn | Controlling |
| Günther, Thomas | 2014 | Planungs- und Kontrollinstrumente zur unternehmenswertorientierten Führung in mittelständischen Unternehmen | Controlling |

| Authors | Year | Title | Publisher |
|---|---|---|---|
| Hartmann, Frank; Maas, Karen; Perego, Paolo | 2016 | Den Wald vor lauter Bäumen nicht sehen: Controller auf der Suche nach Nachhaltigkeit | Springer Gabler |
| Heimel, Jana; Momberg, Martin | 2021 | Sustainable Finance: Nachhaltigkeitscontrolling zur Steuerung des sozialen und ökologischen Wirtschaftens von Unternehmen | Springer Gabler |
| Hoekstra, Rutger | 2020 | SNA and beyond: Towards a Broader Accounting Framework That Links the SNA, SDGs and Other Global Initiatives | Statistical Journal of the IAOS |
| Horváth, Péter; Pütter, Judith M.; Dagilienė, Lina; Dimante, Dzineta; Haldma, Toomas; Kochalski, Cezary; Král, Bohumil; Labaš, Davor; Lääts, Kertu; Bedenik, Nidžara Osmanagić; Pakšiová, Renáta; Petera, Petr; Ratajczak, Piotr; | 2017 | Status Quo and Future Development of Sustainability Reporting in Central and Eastern Europe | Journal for East European Management Studies |

| Krivačić, Dubravka; Janković, Sandra | 2021 | Sustainability reporting during the pandemic: current state and expectations for the future | Journal of Accounting & Management |
| --- | --- | --- | --- |
| Kuttner, Michael | 2022 | Corporate Social Responsibility-Controlling: Eine instrumentelle Perspektive | Springer Nature |
| Lühn, Michael; Nuzum, Anne-Katrin; Petersen, Holger; Schaltegger, Stefan | 2022 | Controller und Nachhaltigkeit im Kontext neuer Berichterstattungspflichten | Rethinking Finance |
| Lux, Wilfried; Olbert-Bock, Sibylle | 2016 | Strategisches Controlling als Teil des Sustainability Performance Managements - auch für KMU | CONTROLLER Magazin |
| Michel, Uwe; Isensee, Johannes; Stehle, Alexander | 2014 | Sustainability Controlling: Planung, Steuerung und Kontrolle der Realisierung der Nachhaltigkeitsstrategie | Springer |
| Mödritscher, Gernot; Wall, Friederike | 2021 | Nachhaltiger Konsum und seine Verankerung im Controlling | Springer |
| Muller, Roger; Veser, Mark | 2020 | The Current State of Nonfinancial Reporting in Switzerland and Beyond | Internal Approaches for Sustainability Assessment in the External Reporting -- A |
| Wördenweber, Martin | 2017 | Nachhaltigkeitsmanagement | Schäffer-Poeschel |